# What Happens After You Both Swipe Right:

# A Statistical Description of Mobile Dating Communications

Jennie Zhang and Taha Yasseri

Oxford Internet Institute, University of Oxford, 1 St Giles OX13JS Oxford, UK

## Abstract

Mobile dating applications (MDAs) have skyrocketed in popularity in the last few years, with popular MDA Tinder alone matching 26 million pairs of users per day. In addition to becoming an influential part of modern dating culture, MDAs facilitate a unique form of mediated communication: dyadic mobile text messages between pairs of users who are not already acquainted. Furthermore, mobile dating has paved the way for analysis of these digital interactions via massive sets of data generated by the instant matching and messaging functions of its many platforms at an unprecedented scale. This paper looks at one of these sets of data: metadata of approximately two million conversations, containing 19 million messages, exchanged between 400,000 heterosexual users on an MDA. Through computational analysis methods, this study offers the very first large scale quantitative depiction of mobile dating as a whole. We report on differences in how heterosexual male and female users communicate with each other on MDAs, differences in behaviors of dyads of varying degrees of social separation, and factors leading to "success"—operationalized by the exchange of phone numbers between a match. For instance, we report that men initiate 79% of conversations--and while about half of the initial messages are responded to, conversations initiated by men are more likely to be reciprocated. We also report that the length of conversations, the waiting times, and the length of messages have fat-tailed distributions. That said, the majority of reciprocated conversations lead to a phone number exchange within the first 20 messages.

## Introduction

Mobile dating applications (MDAs) have evolved from more traditional online dating methods to become an increasingly popular platform to meet romantic partners. Between 2007 to 2009, more new romantic relationships originated online than through any means other than meeting via mutual friends (Finkel et al., 2012). As smartphone usage becomes more pervasive, MDAs are appearing as a natural progression from traditional web-based online dating methods. The defining distinction of MDAs is the ability to locate users and facilitate instant communications between them. Mobility of smartphone applications allows for constant locale change and rapid communications between users, leading to more dynamic and frequent interactions.

Location-based MDAs, which rely on users' geographic locations for matching purposes, have over 91 million users globally (Dredge, 2015). In 2013, not long after major MDA industry giant Tinder (gotinder.co) launched, 3% of the adult American population had used an MDA at one point or another (Smith & Duggan, 2013). As of February 2015, 6% of all Internet users are on MDAs, 62% of which are men and 70% of which are between the ages of 16 and 34 (Dredge, 2015). Tinder reports that it facilitates 26 million matches between users everyday, and that it has seen over 10 billion matches in 196 countries since its launch in the summer of 2012 (Tinder, 2016). The steep upward trend of MDA use has prompted a gap in knowledge about user behavior on these applications.

In response to this gap in knowledge, this work conducts a quantitative examination of text-based communication data on an MDA platform provided by a private company that created and operates the MDA.

Online dating is the practice of using dating sites—made specifically for users to meet each other for the end goal of finding a romantic partner (Finkel et al., 2012). It has an annual growth rate of 70% in the United States (Kaufmann, 2012). At the time of writing, a third of marriages in the U.S. in the last year originated from online dating (Ansari & Klinenberg, 2015). MDAs, for the most part, are a more expedient version of online dating: users mostly can only access the service via their mobile devices.

This expediency is largely due to the speed in which users can converse. An examination of email reply times reveals that the shortest response times come from emails sent from mobile phones (Kooti et al., 2015). In addition to quick reply times, mobile text messaging (MTM) is typically shorter than emails sent from desktops or laptops (Kooti et al., 2015). Dating site OkCupid (okcupid.com) reported a 100-character drop in message lengths immediately after they introduced their mobile application in 2008 (Rudder, 2014). This drop continued until messages were at an average of 100 characters per message, a drop of over two-thirds from the pre-application average (Rudder, 2014). Shorter messages on mobile phones could be due to remnants of older hallmarks of SMS messages, such as character limitations. While many of these limitations do not exist anymore, MTMs, including those sent over MDAs, have retained many of these characteristics. Rudder (2014) found that the fastest response rates occurred when messages were between 40 and 60 characters, meaning that shorter messages led to the most replies. Quicker reply times and shorter messages usually indicate higher rates of frequency in response times between users. These are two characteristics of MTMs, as the immediacy and mobility of devices allow for near-conversational levels of messaging, akin to online instant messaging.

While MTM is faster and more accessible, it still benefits from the "editability" of e-mail and other web-based texts communications (Reid & Reid, 2004). The specific crafting of text provides the dual environment for "intimate personal contact" and the "necessary detachment" to manage self-presentation and involvement (Reid & Reid, 2004). Given that MDA users engage in dyadic communications with people they do not know, self-presentation via messages is key. MDA users can craft the best versions of themselves by taking the time to draft messages, meanwhile still operating within the intimate framework of constant and instantaneous contact. Mediated communication allows for individuals to consciously or subconsciously create a persona that they use to communicate with one another (Tong & Walther, 2012). Given this opportunity, someone could tailor their messages to a specific audience, which, in the case of MDAs, would be their matches.

In this work, we examine data generated by users on a particular MDA (the exact platform must remain anonymous due to the terms of the NDA in place). Like many recently developed MDAs, this platform affords each user with a profile with rudimentary information comprised of a mini biography and pictures. In addition, each user gets access to a certain number of profiles belonging to other users each day. After signing up for the service, users must specify their own gender and the gender(s) of the potential partner. Each user is then shown the profiles of other users of the desired gender(s) and age range within a desired geographical distance. The user then "likes" or "dislikes" each profile by swiping the profile to right and left respectively or pressing a button. If two users mutually like each other's profiles, they are then considered a "match" and can begin conversing with each other. If neither user likes each other, if a like is unreciprocated, or one of the users decides to "unmatch", they never see each other's profiles again. Users are not notified if other users find them desirable previous to a mutual selection, so matches are, theoretically, entirely mutual and made without any existing knowledge of the person's feelings towards them. Ansari and Klinenberg (2015) coin this the "mutual-interest requirement," in which users cannot engage with each other unless they have both indicated some level of interest in each other. We explore what occurs after the match—the dynamics of the conversations that occur between two people who have fulfilled the mutual-interest requirement and where at least one user has attempted to contact the other user.

After matching, a pair of users can begin conversing. The nature of MDA text messages—and modern communication methods—means that rejections will most likely come in the form of silence (Ansari & Klinenberg, 2015). Tong & Walther (2011) report that women are more likely to not respond to date requests than they are to send rejection emails. In online dating as a whole, 26% of men responded to received messages, while women only responded to 16% of total messages (Fiore et al., 2010). Though conversing on MDAs, like most MTMs, is dictated by speed, users can still take the time to craft their messages and present themselves in the best light. Pauses that would have been strange over the phone or face-to-face, tend to be more acceptable in instant messaging or MTM (Whitty et al., 2007). This could also contribute to efforts by users to retain the upper hand in conversations by communicating less frequently than their matches.

The step following reciprocal contact or engaging in mediated communication is typically face-to-face contact (Whitty et al., 2007; Finkel et al., 2012). This is ultimately the aim of location-based MDAs—to connect users face-to-face after they have met on the technology. On traditional online dating sites, most matches report to have met face-to-face within the month if not the week (Whitty & Carr, 2006). To facilitate FTF meetings, users typically have to exchange email addresses or phone numbers. In this respect, phone number exchange is one of the first indicators of a post-MDA relationship, and ultimately represents the success of MDAs in introducing compatible users to one another. As such, this research will operationalize success as the exchange of phone numbers between users.

## Data and Methods

The data consist of a sample of 400,000 unique heterosexual users from 30 of the largest metropolitan areas in the United States. The mean age of users in the sample is 29, and the median age is 27. The gender ratio is close to half. All users are engaging in heterosexual interactions on the MDA. The data includes 2,088,486 conversations encompassing 18,917,884 messages. Conversation, in this case, is defined as any exchange between a pair of users who have showed mutual interest and "matched." The matches leading to each conversation were made on the MDA between late 2013 and April 2015. The last message exchanged in each conversation, as of April 2015, occurred between 1 January 2015 and 22 April 2015. Each message acts as a data point, and each data point is characterized by a set number of qualities, separated into four distinct categories. Each message has 12 variables, as shown in Table 1.

**Table 1**: Data parameters

| Message Identifiers | Match/Sender | Message Content Metadata |
|---|---|---|
| • Conversation ID<br>• Message sequence number | • Degrees of social separation<br>• Conversation initiator or responder<br>• Gender (M/F) | • Time (minutes since match)<br>• Whether a phone number was exchanged<br>• Number of characters<br>• Number of words<br>• Number of lines<br>• Number of exclamation marks<br>• Number of question marks |

For each message we only have counts of words, characters, and lines, which reveal the length of the message. In addition, counts of question marks and exclamation marks could indicate, among other things, questions or excitement. Phone number exchange is returned with a Boolean value: true if there is a continuous string of ten numbers in the message and false if not.[1] There are a few instances of users sending what seem to be multiple identical messages in a row ("repeat" messages), likely due to a glitch in the platform. However, it is impossible to identify and remove all of these messages from the data as we do not have the content of the messages and cannot separate glitches from intentional repeated messages.

The degree of social separation between users in a match is imported from their Facebook accounts, which are used to login to the application. Degree of social separation measures the social distance between two people via the people who connect them, 1 being the closest connection (friends on Facebook) and 4+ being the furthest social distance in our dataset.

---

[1] Due to platform restrictions, conversations are capped at 100 messages. This means that if conversations exceed 100 messages, only the last 100 messages are retained and shown in the data.

Ethics

As mentioned before, the data provides no hints for personal identification. There is absolutely no personal information, such as age or location of each user, so there is no way of knowing if users are participating in multiple conversations. This is mainly to protect privacy. Messages are identified only by the conversations they are in and the order they come in. Nothing is known about the sender of a message other than the gender, degree of social separation with the message recipient, and whether the same sender initiated the conversation. There is no qualitative information regarding message content. There are no real timestamps, just the time between the sent message and when the match occurs, rounded to the nearest five minute increment.

The name of the MDA will not appear in this work. Users of the MDA are informed of data collection and analysis efforts at the time of sign up, when presented with Terms & Conditions and Privacy Policy agreements.

Confidentiality and data transfer agreements were signed both by the company and the University of Oxford to preserve privacy rights for MDA users. The University of Oxford CUREC (Central University Research Ethics Committee) approved all handling of data and research methods (CUREC number: OII-C1A-15-013).

## Results

### Overall Description

Among 18,917,884 messages in 2,088,486 conversations, men sent 56% of all messages,and initiated 79% of conversations. Initiators, or the first person to send a message in a conversation, sent 54% of all messages. However, of *non*-initial messages, initiators sent 51% of messages. This suggests that though initiators maintain a slight level of dominance in conversations, initiator to non-initiator message ratios tend to balance out as conversations progress.

The length of a message in character follows a fat-tailed distribution with mean of 59.4 and standard deviation of 59.8. However, when logarithmically transformed, the distribution, as seen in Figure 1, fits a lognormal distribution with a peak at 42 characters per message. It also illustrates that a high concentration of messages contain between 30 and 60 characters. The length of messages in words ($M$ = 11.6, $SD$ = 11.5), however, shows a distribution heavier on the left side (Figure 1). The highest peak shows a large concentration of messages around 10 words when the median is 8 words. 37% of all messages have question marks in them and 21% contain exclamation marks. 1.4% of all messages contain a phone number in them.

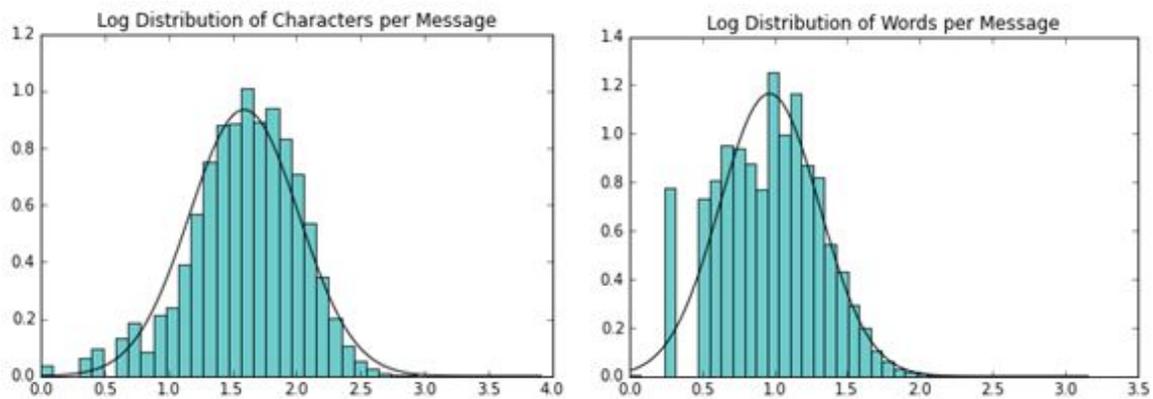

**Figure 1**: Histograms of the logarithm of characters and words per message

89.4% of all reciprocal conversations have a response after just one initial message ($M$ = 1.14, $SD$ = 0.69). This first message occurred, on average, 6.8 days after the match ($SD$ = 4 weeks, mode = < 5 minutes after match). The average first message contains 8 words or 42 characters. About 47% of first messages contain one question mark, 46% contain none, and the rest contain at least two question marks. 70% had no exclamation marks, 25% had one, and 5% had at least two exclamation marks. Of all first messages, 79.4% were sent by men. Of reciprocated first messages, 83% were sent by men. Out of the 2,088,486 first messages, only 1,722 contained phone numbers (about less than 0.01%).

*Conversation Lengths*

2,088,486 conversations comprised of 18,917,884 million messages equates to roughly 9 messages per conversation on average . However, this value does not give an accurate description; 39% of conversations contain just one, unreciprocated message. A further 11% of conversations contain just two messages. This essentially means that only half of all conversations have over two messages. As such, the distribution of messages is quite skewed (see Figure 2).

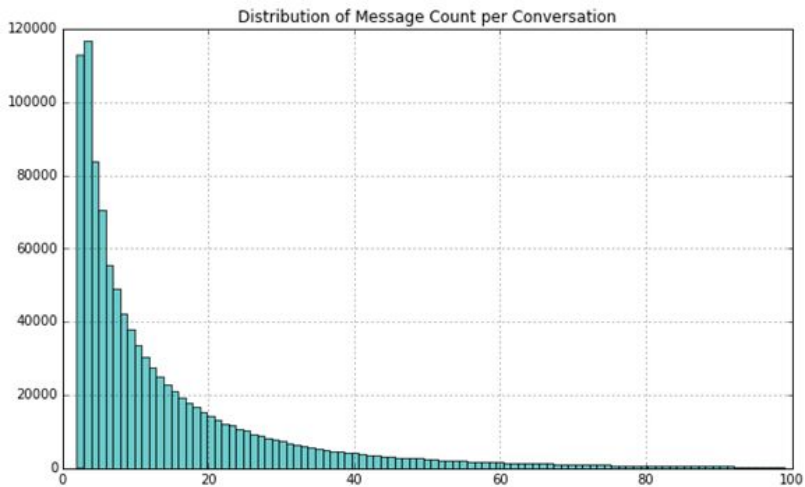

**Figure 2**: Histogram of message count per conversation

While examining the entire dataset gives us a good idea of how often messages are responded to, the results this study is most interested in are from conversations where both users are participants. In addition to the 39% of conversations that consist of only one message, a further 10% contain two or more messages that are unreciprocated. That leaves 1,064,537 conversations (51%) where initiators receive a response, called "mutual conversations" from this point forward. There could be a variety of reasons for unreciprocated matches: recipients could be inactive users, uninterested in the other user, or unimpressed by the initial message. As such, the bulk of our following analysis is on mutual conversations. Mutual conversations encompass 16,983,735 messages, or 90% of the total messages from the original dataset.

After cleaning out unreciprocated, the distribution of messages per conversations remains skewed to the left ($M$ = 14.6, $SD$ = 16.2, median = 8). The most frequent observation, or the mode, of the dataset (11.1% of conversations) is 3 messages per conversation, followed by 2 messages (10.8%).

*Timing*

A large percentage of first messages (15%) occurs right after the match, at 0th minute. In half of the mutual conversations, the first message is sent within 8 hours from the match time with an average of 4.2 days and $SD$ of = 19 days (Figure 3).

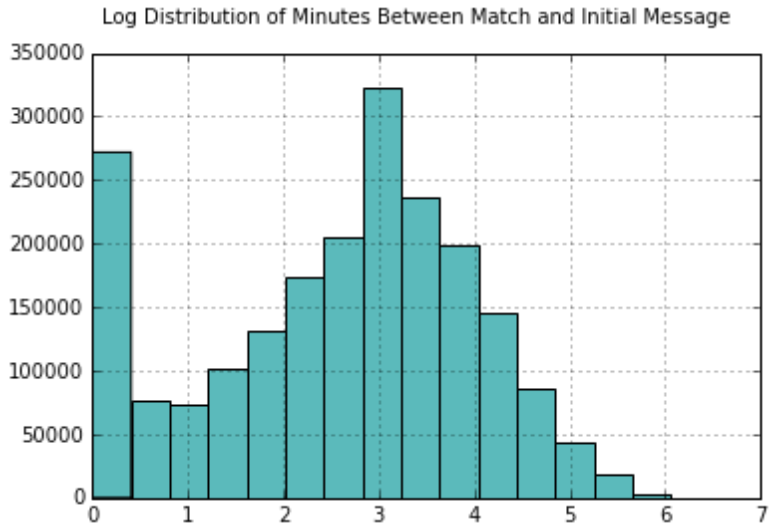

**Figure 3**: Histogram of the logarithm of the time of the first message (minutes) added to 1

The average time it takes between first response and first message sent by initiator is 3,462 minutes or 2.4 days (*SD* = 15.2 days) though the majority of responses still tend to come in within the first few hours of the initial message (see Figure 4).

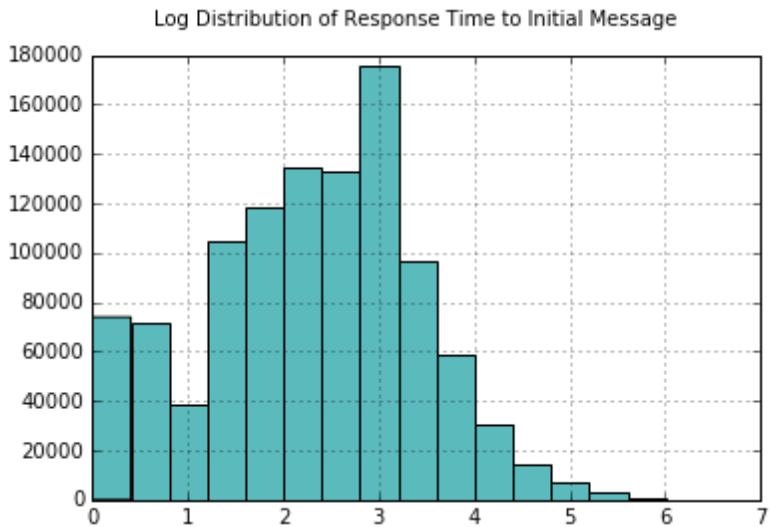

**Figure 4**: Histogram of the logarithm of the response time to the first message (minutes) added to 1

The length of conversations in minutes, defined as the time elapsed between the first and last messages, has a very different distribution to length by message count: instead of a steady decline, length in minutes actually shows a distribution that is close to lognormal (Figure 5, left panel). The median of the distribution is 3,650 minutes, or 2.5 days, and the mode is actually just 5 minutes—despite and the mean is 11 days.

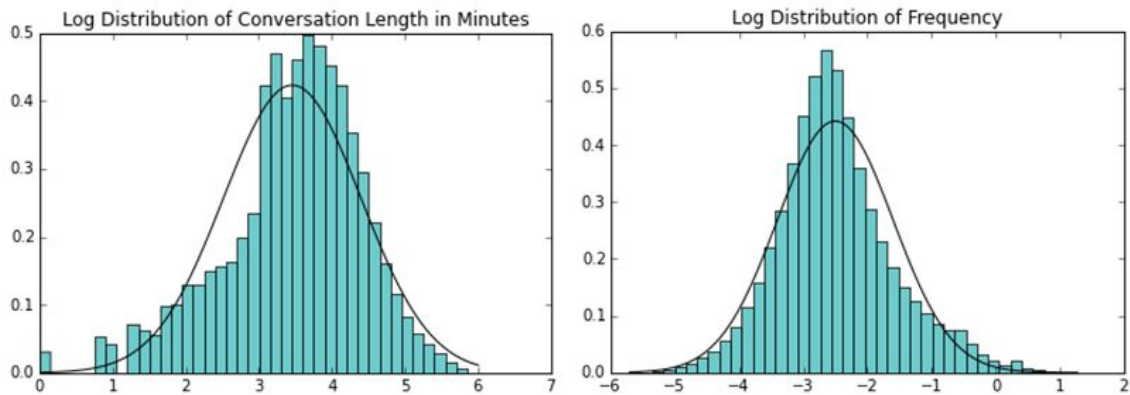

**Figure 5:** Histogram of the logarithmic of, left: conversation lengths in minutes and right: the average message frequency (message per day) for each conversation. The solid line shows a log-normal fit.

The distribution shown in the right panel of Figure 5 is of messages per day in each conversation. Though many conversations occur over very short time periods, converting conversation lengths into days allows for a projection of how many messages would be sent if a match spent all day messaging each other at their average messaging rate. The distribution is fitted with a lognormal—with a prominent peak between the 10 and 100 messages per day marks. The median is 3.8 messages per day.

*Message Lengths*

An overview of number of lines per conversation shows a very similar distribution to number of messages per conversation—this is because the overwhelming majority of all messages (98.3%) only have one line, quite possibly due to the short, quick nature of MTMs (Igarashi et al., 2005). Aside from average lengths of words, which seem mainly concentrated around 5 characters per word, the rest of the conversation or message lengths are mainly distributed log-normally (see Table 2 and Figure 6). This is in accordance with similar observations in online discussion forums (Sobkowicz et al., 2013) and Wikipedia articles (Yasseri et al., 2012).

**Table 2:** Character, message, and word lengths

|  | Total Number of Characters | Total Number of Words | Avg. Characters per Message | Avg. Words per Message | Avg. Lengths of Words |
|---|---|---|---|---|---|
| **Mean** | 896.89 | 174.86 | 58.7 | 11.4 | 5.1 |
| **Median** | 457 | 89 | 50.5 | 9.9 | 5.1 |
| **Mode** | 82 | 11 | 32 | 6 | 5 |
| **logged Mean** | 2.63 | 1.93 | 1.68 | 0.98 | 0.71 |
| **logged SD** | 0.57 | 0.57 | 0.28 | 0.27 | 0.04 |

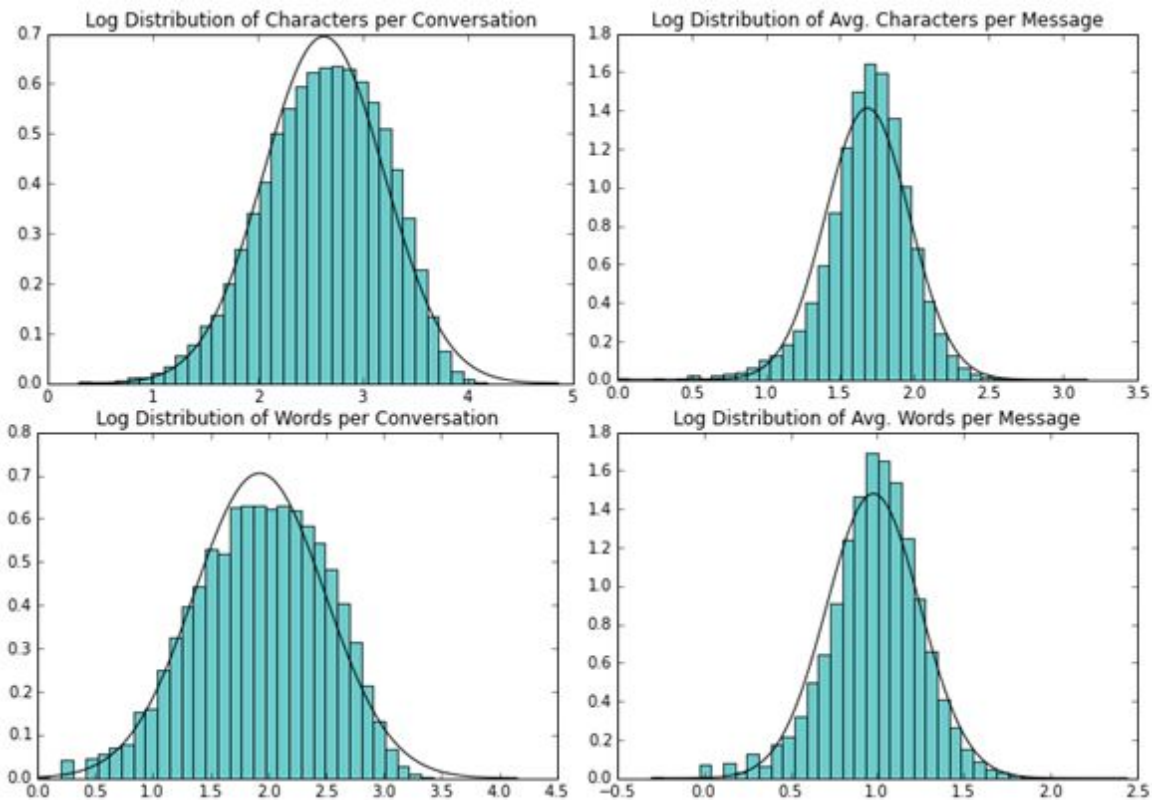

**Figure 6:** Logarithmic histograms of words and characters per conversations and messages

*Gender*

Despite the fact that users in the dataset are approximately half men and half women, the actual message makeup in 56% from men and 44% from women. When unreciprocated messages are cleaned out, the gender makeup of message senders is 54% male and 46% female. Despite a more balanced ratio of message senders, the ratio of male to female initiators actually becomes more imbalanced in mutual conversations, with 83% of initiators male and 17% female compared to 80% male initiators and 20% female initiators in all messages. This highlights a disparity in reciprocity rates: only 42% of the messages sent by female initiators are responded to. Messages sent by male initiators, however, have a 53% response rate.

This is a phenomenon which mimics findings in a 2006 study by Whitty and Carr who found that 60% of men found a particular online dating platform to be a numbers game. While not identical to the MDA matching process, the two platforms share the idea that conversations often begin with the knowledge that both users are interested in each other. Given the number of profiles available, individuals could keep trying until they get a response, meaning they are not fully interested in some of the profiles they send "kisses" to. Instead, they would send a large number of kisses and see which women reciprocate (Whitty & Carr, 2006). MDAs provide an ideal platform for men who practice this and leads to a potential imbalance: users can swipe right on all the profiles they see and then filter based on who likes them back.

*Punctuation Usage*

There are some pointed differences in how men and women use punctuation on MDAs. Men are more likely to send questions marks while women are more likely to send exclamation marks (see Table 3). For both men and women, a significant percentage of messages with exclamation marks, have more than one mark in them (17% and 20% respectively). Similarly, among those with question marks, there are many messages containing more than one questions marks in them (13% for men and 11% for women).

**Table 3:** Percentages of punctuation use in messages by gender

|  | **Messages sent by men** | **Messages sent by women** | **All Messages** |
|---|---|---|---|
| **Exclamation Marks** | 17% | 26% | 21% |
| **Question Marks** | 40.5% | 33.5% | 37% |

To investigate punctuation use, we compare ratios of punctuation between men and women at the level of individual conversations. Due to varying lengths of conversations, ratios allow for comparison across all conversations. Ratios are calculated by placing the female value over the value of the sum. Ratios in which neither men or women use the specific punctuation mark (i.e. 0 over 0) have been excluded from the calculation. Conversations without exclamation marks made up 24% of mutual conversations, while 7% of conversations did not have question marks.

As seen in the first two columns of Table 4, low mean and median values of question mark ratio show a tendency for men to use questions marks at a slightly higher rate than women. However, the mode is still 0.5, meaning the most common occurrence is men and women using them at the same rate. Exclamation mark use shows a larger imbalance, with women using more exclamation marks. This is also expressed in the mode of 1, showing that the most common conversation is one where the woman is the sole user of exclamation marks.

**Table 4:** Statistics of conversation ratios

|  | **Question Mark Ratio** | **Exclamation Mark Ratio** | **Message Ratio** | **Word Ratio** | **Character Ratio** |
|---|---|---|---|---|---|
| **Mean** | 0.41 | 0.60 | 0.43 | 0.45 | 0.44 |
| **SD** | 0.27 | 0.36 | 0.11 | 0.16 | 0.17 |
| **Median** | 0.4 | 0.6 | 0.44 | 0.45 | 0.45 |
| **Mode** | 0.5 | 1 | 0.5 | 0.5 | 0.5 |

As seen in the three right columns of Table 4, all ratios depicting conversation lengths show heavier male usage. However modes are all 0.5—given that a large percentage (10.6%) of mutual conversations consist of just two messages (one from each user), this is self-evident. The second most common message ratio is 0.33, most likely from conversations of three messages (two sent by the man and one sent by the woman). Other than that, lower means and medians are in keeping with overall depiction of men sending more messages than women. On average, women send 7 messages in a conversation (*SD* = 8, median = 4) and men send 8 messages (*SD* = 9, median = 5). Word and characters per message are almost identical across genders, with men having slightly more to say per message—average word per message is identical to overall mean (11.6 words) and average character per message is slightly higher (59.6 characters compared to 59.4). These values for women are 11.5 words and 59.0 characters.

When conversations initiated by men are isolated, the ratios within these conversations show few differences from all mutual conversations (see Table 5). Ratios are lower, showing a slight dominance by men in terms of punctuation and conversation lengths. However, when conversations initiated by females are isolated, distributions of ratios are mirrored. Figure 7 exemplifies this in message ratio. Both distributions of message ratio peak at 0.5, but the male-initiated conversations have more instances of values to the left—the second highest peak occurs between 0.3 and 0.4. Female-initiated conversations show the opposite: the second highest peak occurs between 0.6 and 0.7. This is true for most other ratios (see Table 5), except for exclamation mark ratio, in which case female-initiated conversations show a greater imbalance of women using more exclamation marks than men. The mirrored distributions in female-initiated conversations and increased ratio values could indicate that initiator status, rather than gender, may have more of an impact on a user's participation rate.

**Table 5:** Statistics of conversation ratios by gender of initiator

|  | Male | | | Female | | |
|---|---|---|---|---|---|---|
|  | **Mean** | **SD** | **Median** | **Mean** | **SD** | **Median** |
| **Exclamation Mark Ratio** | 0.59 | 0.36 | 0.60 | 0.64 | 0.35 | 0.67 |
| **Question Mark Ratio** | 0.39 | 0.27 | 0.40 | 0.52 | 0.28 | 0.50 |
| **Message Ratio** | 0.43 | 0.10 | 0.44 | 0.54 | 0.11 | 0.50 |
| **Word Ratio** | 0.43 | 0.16 | 0.43 | 0.52 | 0.17 | 0.52 |
| **Character Ratio** | 0.43 | 0.16 | 0.43 | 0.52 | 0.18 | 0.52 |

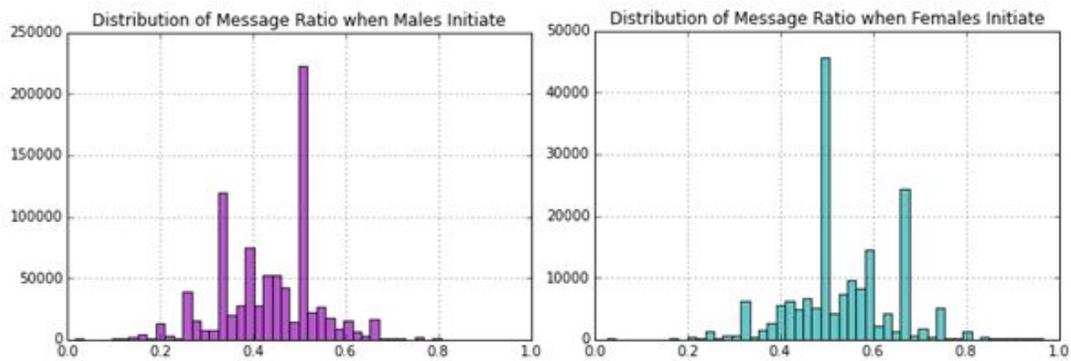

**Figure 7:** Histograms of message ratios separated for when males or females initiate

## Success Rate

Oftentimes, a phone number exchange indicates that the conversing pair is ready to move the conversation from the MDA to another platform as a transition to the next step of their online relationship. Mutual conversations with phone number exchanges will be operationalized as "successful" as it signifies at least one user revealing their personal information to the other user, thus indicating that the MDA has served its purpose in introducing a pair of users to each other. 19% of mutual conversations include at least one number exchange and 17.3% contained phone numbers sent by both parties. Among those conversations that have only one party's phone number, females were the sole phone sharers in 57.3% of cases.

The mean message in which phone number is exchanged is the 27th message ($SD$ = 20, median = 22, mode = 12). The phone number is typically exchanged towards the end of the conversation. The relative phone number exchange position per conversation (message sequence number divided by total message count), has a mean of 0.94 ($SD$ = 0.13), indicating that the average phone number exchange occurs when the conversation is 94% over. The median and mode are both 1, signifying that phone numbers are most commonly exchanged on the last message of the conversation.

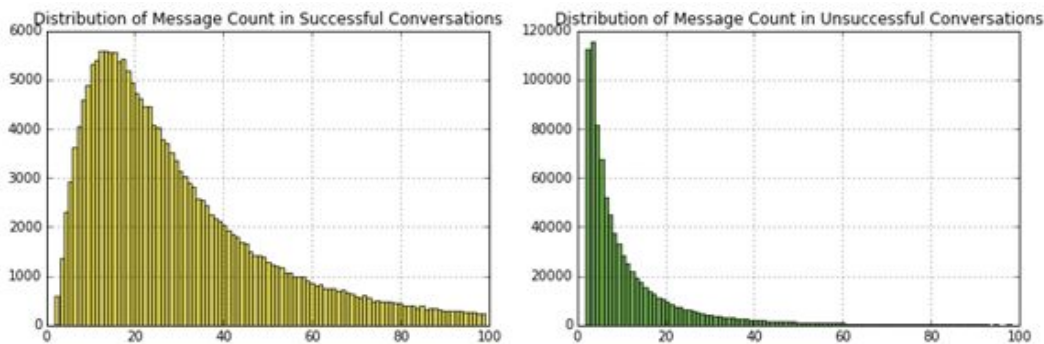

**Figure 8:** Histograms of successful message count and unsuccessful message count

One of the most distinct differences between successful and unsuccessful conversations can be seen in message count distributions (see Figure 8). Successful conversations feature a peak between 10 and 20 messages with an extended right tail ($M$ = 29, $SD$ = 20, median = 23, mode = 3). Unsuccessful conversations, on the other hand peak at the lowest values and drop steeply down ($M$ = 11, $SD$ = 13, median = 6, mode = 3).

The lengths in time of the two types of conversations show similar patterns: successful conversations ($M$ = 14.2 days, $SD$ = 32.5 days, median = 4.5 days) tend to last longer than unsuccessful conversations ($M$ = 10.2 days, $SD$ = 28.2 days, median = 2.1 days). The most common length of time, or mode, in successful conversations and unsuccessful conversations is 25 minutes and 5 minutes, respectively.

**Table 6:** Comparison of message and conversation lengths in successful and unsuccessful conversations

|  | All | | | Successful | | | Unsuccessful | | |
|---|---|---|---|---|---|---|---|---|---|
|  | Mean | Mode | Median | Mean | Mode | Median | Mean | Mode | Median |
| **Character Count** | 897 | 82 | 457 | 1,983 | 806 | 1,556 | 654 | 896 | 338 |
| **Character Ratio** | 0.44 | 0.5 | 0.45 | 0.46 | 0.5 | 0.46 | 0.44 | 0.5 | 0.44 |
| **Word Count** | 175 | 11 | 89 | 386 | 141 | 304 | 128 | 11 | 66 |
| **Word Ratio** | 0.45 | 0.5 | 0.45 | 0.46 | 0.5 | 0.46 | 0.44 | 0.5 | 0.44 |
| **Message Count** | 15 | 3 | 8 | 29 | 12 | 23 | 11 | 3 | 6 |
| **Message Ratio** | 0.45 | 0.5 | 0.47 | 0.47 | 0.5 | 0.47 | 0.45 | 0.5 | 0.46 |
| **Question Mark Count** | 6.3 | 4 | 2 | 12 | 8 | 10 | 5 | 2 | 3 |
| **Question Mark Ratio** | 0.41 | 0.4 | 0.50 | 0.41 | 0.41 | 0.50 | 0.41 | 0.4 | 0.50 |
| **Exclamation Mark Count** | 3.9 | 0 | 2 | 8 | 1 | 5 | 3 | 0 | 2 |
| **Exclamation Mark Ratio** | 0.6 | 0.6 | 0.50 | 0.61 | 0.63 | 0.50 | 0.59 | 0.6 | 0.50 |

The message count sent by the initiator over the total message count within a conversation, is slightly different across the two types of conversations. Successful conversations have a mean initiator ratio (percentage of messages sent by the initiator) of 0.54 ($SD$ = 0.09) and unsuccessful conversations have a mean ratio of 0.57 ($SD$ = 0.11). This indicates that successful conversations have a more balanced message count between initiators and recipients.

Table 6 shows how all mutual conversations compare with successful and unsuccessful conversations by aligning statistics of all counts and gender ratios of messages, words and characters. There are some drastic differences: successful conversations tend to have higher counts than unsuccessful conversations across the board. In addition, successful conversations show higher uses of question and exclamation marks indicating a better crafted messaging. Gender ratios are more balanced in successful conversations in all categories except for exclamation ratio—successful conversations show an even higher imbalance of women using exclamation marks more than men.

*Predicting Success*

The above analysis led to the identification of variables to regress with phone number exchange. All variables are attributes of mutual conversations and we do not have any prior information on the parties involved. To test the relationship between each of these independent variables and success, individual logistic regressions were conducted on each variable and whether or not a phone number was exchanged.

19% of mutual conversations contain phone numbers and 81% do not—as such, the overall odds of a successful conversation are 19 to 81, or 0.235. The results from the individual logistic regressions of each continuous variable are in Table 7. The two values reported per variable are the odds ratio and the R-squared value. None of the variables result in very strong odds ratios on their own. Most waver around 1—though the most notable exception is message ratio, with an odds ratio of 5.6. Odds ratios are above 1 and indicate a positive relationship, meaning that an increase in the independent variable results in the increased odds of a successful conversation. For ratio variables, this indicates increased women to men usage of the variable would be more likely to end in success. The initiator ratio shows that the less an initiator participates in relation to the recipient, the better chances of a successful conversation. Many of the individual models have very small R-squared. However, there are a few exceptions: message count, female question mark count, and male exclamation mark count.

**Table 7:** Individual logistic regression table

| Variable | Odds Ratio | R-Squared |
| --- | --- | --- |
| **Frequency** | 1.0001 | 0.0005 |
| **Initiator Message Ratio** | 0.7447 | 0.0115 |
| **Message Count** | 1.0586 | 0.1548 |
| **Message Count (Women)** | 1.0090 | 0.1571 |
| **Message Count (Men)** | 1.0084 | 0.1652 |
| **Message Ratio (Woman/Man)** | 5.5894 | 0.0058 |
| **Word Count** | 1.0046 | 0.1714 |
| **Word Count (Women)** | 1.0090 | 0.1571 |
| **Word Count (Men)** | 1.0084 | 0.1652 |
| **Word Ratio (Woman/Man)** | 1.5993 | 0.0009 |
| **Character Count** | 1.0009 | 0.1710 |
| **Character Count (Women)** | 1.0018 | 0.1568 |
| **Character Count (Men)** | 1.0016 | 0.1644 |
| **Character Ratio (Woman/Man)** | 1.6389 | 0.0011 |
| **Question Mark Count** | 1.1547 | 0.1538 |
| **Question Mark Count (Women)** | 1.2893 | 0.1172 |
| **Question Mark Count (Men)** | 1.2425 | 0.1344 |
| **Question Ratio (Woman/Man)** | 0.9665 | 0.0000 |
| **Exclamation Mark Count** | 1.1244 | 0.0856 |
| **Exclamation Mark Count (Women)** | 1.1887 | 0.0756 |
| **Exclamation Mark Count (Men)** | 1.1837 | 0.0509 |
| **Exclamation Ratio (Woman/Man)** | 1.3455 | 0.0014 |

## Degrees of Separation

Each message in the dataset is assigned a "degree of social separation" which indicates how closely tied the message sender is to the recipient. Table 8 includes frequencies of each degree and the percentage makeup of each in three types of conversations: all, mutual, and successful. The majority of conversations (59%) are 3-degrees, followed by 2-degree (34%) and +4 (7%) conversations. There is a small minority of conversations that have a degree of separation of 1.

The chance of reciprocation is significantly higher for 2-degree and 3-degree conversations compared to +4-degree. However, considering the rate of phone number exchange per reciprocated conversation, 3-degree conversations are similar to +4-degree both slightly less than for 2-degree. This shows that even though the reciprocation rate varies with the degree of separation, as soon as the conversation is reciprocated, the effect of social separation vanishes. The rate of phone number exchange per conversation (regardless of reciprocation), decreases by the degree of separation monotonically. The case of 1-degree separation is very special, those pairs are already facebook friends, even though the rate of reciprocation is very high, the phone number exchange is not very common.

**Table 8:** Breakdown of conversations by degree of social separation

| Degree of Separation | Number of Conversations | Rate of Reciprocation | Rate of Phone Number Exchange per conversation | Rate of Phone Number Exchange per Mutual Conversation |
|---|---|---|---|---|
| 1 | 432 (0.02%) | 0.75 | 0.07 | 0.09 |
| 2 | 709,268 (34%) | 0.54 | 0.11 | 0.20 |
| 3 | 1,236,994 (59%) | 0.52 | 0.09 | 0.18 |
| +4 | 141,792 (7%) | 0.26 | 0.05 | 0.18 |

## Discussion and Conclusions

The results provide an extensive quantitative depiction of how heterosexual MDA users communicate with one another. In addition to general findings, this quantitative depiction encompasses differences in gender behavior on MDAs, potential differences between matches of various degrees of separation, and differences between successful and unsuccessful conversations.

Almost half (49%) of all MDA conversations consist of unreciprocated messages sent by one user. This confirms previous findings that date requests made on mediated form of communications are less likely to be responded to (Tong & Walther, 2011). In the dataset of reciprocated conversations, the most common lengths are 3 and 2 messages, respectively. Less than 1% of these messages result in phone number exchanges, so a good number of the responses in these conversations could very well be considered as rejection messages or indicate a drop in interest.

A good percentage of conversations actually occur over periods of over a day to a week. However, the large majority of conversations do not last over three to four weeks. In this respect, MDAs could very likely emulate most online dating platforms where most users meet within the month—if not the week—and if arrangements do not occur in this time period, users most likely will not continue talking or meet at all (Rosen et al., 2008). Though conversations most frequently begin within 5 minutes of a match, the distribution has more of a spread, meaning that start times are quite varied. That said, most conversations (71%) do begin within a week of a match. The nature of MDA profiles and how quickly users can view profiles before making a judgment could contribute to the speed between match and message: it is likely that users are already wholly familiar with a profile before initiating a conversation.

Average frequencies of conversations show that despite what might be a speedier platform, messages are still fairly spread out, with a median of about four messages a day. There are a large amount of conversations with messages that happen over a short period of time (the mode is about a message every three minutes) but, as average frequencies do not take the large differences in intervals between messages, this is not a variable that can accurately depict frequencies in MDA conversations. This suggests that the communication happens in bursts of messages followed by long silence period which has been reported in many different digital platforms (Vázquez et al., 2006).

Average message lengths (59 to 60 characters) on the MDA are smaller than message lengths on traditional online dating sites—OkCupid founder Christian Rudder (2014) reports a 100-character average per message. The higher character length on OkCupid is most likely due to the fact that many users still access the platform on a PC rather than their mobile phones. The shorter messages found on the MDA examined here confirm Igarashi et al.'s (2005) notion that MTMs are shorter and quicker and Rudder's notion (2014) that their mobile application was the reason for the steep drop in character-average per message.

One of the most drastic differences between men and women is how often they initiate conversations. Messages are five times more likely to have been initiated by a man than by a woman, which confirms previous work that found men to be the main initiators in heterosexual conversations (Finkel et al., 2012; Whitty, 2012; Tong & Walther, 2011; Whitty et al., 2007). It is worth noting that reciprocation rates differ than previous findings, which found that in addition to initiating, men were more likely to respond—even if it was in a form of rejection (Tong & Walther, 2011; Fiore et al., 2010). In our dataset however, only 42% of messages sent by women were responded to while 53% of messages sent by men were responded to. This could be due to a variety of reasons. The first is that men could be behaving on the MDA similarly to how they behaved on the Australian site that Whitty and Carr (2006) observed—by treating the platform as a "numbers game," men could increase chances of matching and "like" as many profiles as possible. They could then react selectively after seeing which women reciprocated. This could result in male users matching with more profiles than they are interested in and choosing to ignore matches they later decide they are no longer interested in.

Other marked differences include how men and women use punctuation. Out of all the ratios, the ratio of women using exclamation marks to men using exclamation marks is the highest, which matches Fox et. al.'s (2007) findings that women are more expressive in their messages. Question marks, on the other hand, are used more frequently than by men. On that note, men are also found to have sent less phone numbers which could indicate that they are the requesters of numbers.

Despite previous findings that men received longer messages than women did (Fox et al., 2007), in our dataset, men tend to send slightly longer messages and slightly more messages in a conversation than their female counterparts. However, the ratios of female usage of messages, words, and characters to overall usage is 0.45, showing that despite a slight skew towards males, the ratios are still fairly balanced. In addition, when conversations initiated by females are isolated, these ratios are flipped and women are actually the sender of longer messages, more messages, and more questions marks. This shows that female messaging behavior is fairly different when they initiate and behavior is more tied to initiation than gender.

Phone number exchange makes a good operationalized outcome. In most conversations, it is typically exchanged in the last few messages of a conversation before a match presumably changes platforms via the phone number. However, it is important to note that any differences found in successful and unsuccessful conversations are not necessarily indicators of factors that may have led to one outcome or another. While there might be apparent differences between the two types of conversations, they are not necessarily causes for a particular outcome.

One of the most blatant differences between successful and unsuccessful conversations is the distribution of number of messages per conversation. Successful conversations tend to contain many more messages and are longer in length time-wise. Ratios of messaging and conversation lengths were also more balanced—in both gender and initiator status. In addition, successful conversations show both participants using more question and exclamation marks. Only 1% of successful conversations do not contain question marks and only 9% do not contain exclamation marks—the respective proportions in unsuccessful conversations are 8% and 28%.

Regressing individual variables against phone number exchange as a binary dependent variable result in message count, female question mark count, and male exclamation mark count as variables with the strongest predictive power. In addition, initiator ratio evinces a negative relationship with success, meaning that more recipient participation in relation to initiator participation could indicate a successful conversation.

MDAs that include information about mutual friends are essentially perpetuating the idea that people are interested in knowing how they are connected to their potential romantic partners. Since MDAs typically connect users to those they do not know in real life, they are widening the social circles of users for dating purposes. This emulates the real-world actions of attempting to meet "weak ties"—or friends of friends and other acquaintances (Christakis & Fowler, 2009). By providing users with profiles of people they do not know and revealing information on potential shared friends or friends of friends, many MDAs are connecting weak ties with one another. Christakis and Fowler (2009) emphasize the appeal of weak ties in dating for simultaneously providing expanded pools of options and for creating common ground between potential partners.

This could explain the tendency for pairs with no apparent ties to engage in mutual conversations at much lower rates than pairs of closer ties. In addition, this confirms Tong and Walther's finding (2011) that people are less likely to reject weak ties if there is a chance of encountering them in the future. Though the majority of conversations on this MDA are between matches with a degree of social separation of 3, the conversion rate of unreciprocated to reciprocated conversation and mutual conversation to successful conversation is highest in conversations of a degree of 2. This is followed by 3-degree and +4-degree conversations, respectively.

However, once conversations are reciprocated, the percentages of successful results tend to even out: null-degree and 3-degree conversations have almost the same rates of phone number exchange. This could show that at a certain threshold of conversation or common ground, matches of non-existent and of weak ties are equally likely to exchange phone numbers with each other. Previous research shows that marriages where spouses are the sole connectors for their respective friend groups are actually stronger than marriages in which spouses have highly interweaved friend groups (Backstrom & Kleinberg, 2014), indicating that ties might have more of an impact on whether two people decide to meet rather than the outcome of their potential relationship.

Here we presented, for the first time, a large scale quantitative description of the ever growing MDA phenomenon. In short, we reported that a typical MDA conversation lasts an average length of 11 days and 15 messages, with each message possessing an average of 11 words. A conversation will typically have around 6 question marks and 4 exclamation marks. Phone numbers are, on average, exchanged around the 27th message. Phone numbers typically mark the end of the conversation on the MDA, as it is assumed that users will continue conversing over text message. Phone number exchange happens in about 19% of mutual conversations and 1.4% of all conversations. Successful conversations average about 30 messages versus 11 messages in unsuccessful conversations. There are many aspects of mobile dating and communicating on MDA platforms that have yet to be explored. Although some of this is due to the limitations of the dataset, much of it is simply due to the fact that MDAs are very new and there are still an abundance of features to examine.

## Acknowledgments

We thank Ralph Schroeder and Grant Blank for their comments and suggestions.

## References

Ansari, A., & Klinenberg, E. (2015). *Modern Romance*. Penguin Publishing Group.

Backstrom, L., & Kleinberg, J. (2014). Romantic partnerships and the dispersion of social ties. In *Proceedings of the 17th ACM conference on Computer supported cooperative work & social computing - CSCW '14* (pp. 831–841). Physics and Society, New York, New York, USA: ACM Press. doi:10.1145/2531602.2531642

Christakis, N. A., & Fowler, J. H. (2009). *Connected: The Surprising Power of Our Social Networks and How They Shape Our Lives*. New York (Vol. 3). doi:10.1111/j.1756-2589.2011.00097.x

Dredge, S. (2015, February 17). Nearly two thirds of mobile dating app users are men. *The Guardian*. Retrieved from

http://www.theguardian.com/technology/2015/feb/17/mobile-dating-apps-tinder-two-thirds-men

Finkel, E. J., Eastwick, P. W., Karney, B. R., Reis, H. T., & Sprecher, S. (2012). Online Dating: A Critical Analysis From the Perspective of Psychological Science. *Psychological Science in the Public Interest*, *13*(1), 3–66. doi:10.1177/1529100612436522

Fiore, A. T., Taylor, L. S., Zhong, X., Mendelsohn, G. a., & Cheshire, C. (2010). Who's right and who writes: People, profiles, contacts, and replies in online dating. *Proceedings of the Annual Hawaii International Conference on System Sciences*, 1–10. doi:10.1109/HICSS.2010.444

Fox, A. B., Bukatko, D., Hallahan, M., & Crawford, M. (2007). The Medium Makes a Difference: Gender Similarities and Differences in Instant Messaging. *Journal of Language and Social Psychology*, *26*(4), 389–397. doi:10.1177/0261927X07306982

Igarashi, T., Takai, J., & Yoshida, T. (2005). Gender differences in social network development via mobile phone text messages: A longitudinal study. *Journal of Social and Personal Relationships* , *22* (5 ), 691–713. doi:10.1177/0265407505056492

Kaufmann, J.-C. (2012). *Love Online*. Polity.

Kooti, F., Aiello, L. M., & Lerman, K. (2015). Evolution of Conversations in the Age of Email Overload. *WWW*.

Reid, F. J. M., & Reid, D. J. (2004). Text appeal: The psychology of SMS texting and its implications for the design of mobile phone interfaces. *Campus - Wide Information Systems*, *21*(5), 196–200.

Rosen, L. D., Cheever, N. A., Cummings, C., & Felt, J. (2008). The impact of emotionality and self-disclosure on online dating versus traditional dating. *Computers in Human Behavior*, *24*(5), 2124–2157. doi:10.1016/j.chb.2007.10.003

Rudder, C. (2014). *Dataclysm: Who We Are (When We Think No One's Looking)*. Random House of Canada.

Smith, A., & Duggan, M. (2013). *Online dating & relationships*. *Pew Internet & American Life Project*. Retrieved from http://pewinternet.org/~/media/Files/Reports/2013/PIP_Online Dating 2013.pdf

Sobkowicz, P., Thelwall, M., Buckley, K., Paltoglou, G., & Sobkowicz, A. (2013). Lognormal distributions of user post lengths in Internet discussions-a consequence of the Weber-Fechner law? *EPJ Data Science*, 2(1), 1.

Tinder. (2016). Tinder One Sheet.

Tong, S. T., & Walther, J. B. (2011). "Just say no thanks": Romantic rejection in computer-mediated communication. *Journal of Social and Personal Relationships*, *28*(4), 488–506. doi:10.1177/0265407510384895

Tong, S. T., & Walther, J. B. (2012). The Confirmation and Disconfirmation of Expectancies in Computer-Mediated Communication. *Communication Research*, 0093650212466257–. doi:10.1177/0093650212466257

Vázquez, A., Oliveira, J. G., Dezsö, Z., Goh, K. I., Kondor, I., & Barabási, A. L. (2006). Modeling bursts and heavy tails in human dynamics. *Physical Review E*, 73(3), 036127.

Whitty, M. T., Baker, A. J., & Inman, J. A. (2007). *Online Matchmaking* (Vol. 0). Palgrave Macmillan.

Whitty, M. T., & Carr, A. N. (2006). *Cyberspace Romance: The Psychology of Online Relationships*. Palgrave Macmillan.

Yasseri, T., Sumi, R., Rung, A., Kornai, A., & Kertész, J. (2012). Dynamics of conflicts in Wikipedia. *PloS one*, 7(6), e38869.